\documentclass[10pt]{iopart}
\usepackage{graphicx}
\usepackage{caption}
\usepackage[compatibility=false]{caption}

\begin{document}

\title[]{Effect of interface geometry on electron tunnelling in Al/Al$_2$O$_3$/Al junctions}

\author{M Koberidze$^1$, A V Feshchenko$^2$, M J Puska$^1$, R M Nieminen$^1$ and J P Pekola$^2$} 
\address{$^1$ COMP Centre of Excellence, Department of Applied Physics,  Aalto University School of Science, P.O. Box 11100, FI-00076,  Aalto, Finland}
\address{$^2$ Low Temperature Laboratory, Department of Applied Physics, Aalto University School of Science, P.O. Box 15100, FI-00076 Aalto, Finland}

\ead{manana.koberidze@aalto.fi}

\begin{abstract}
We investigate how different interface geometries of an Al/Al$_2$O$_3$ junction, a common component of modern tunnel devices, affect electron transport through the tunnel barrier. We study six distinct Al/Al$_2$O$_3$ interfaces which differ in stacking sequences of the metal and the oxide surface atoms and the oxide termination. To construct model potential barrier profiles for each examined geometry, we rely on first-principles density-functional theory (DFT) calculations for the barrier heights and the shapes of the interface regions as well as on experimental data for the barrier widths. We show that even tiny variations in the atomic arrangement at the interface cause significant changes in the tunnel barrier parameters and, consequently, in electron transport properties. Especially, we obtain that variations in the crucial barrier heights and widths can be as large as 2 eV and 5 \AA, respectively. Finally, to gain information about the average properties of the measured junction, we fit the conductance calculated within the WKB approximation to the experimental data and interpret the fit parameters with the help of the DFT results.
\end{abstract}

\pacs{73.40.Gk, 73.40.Rw, 73.30.+y}

\vspace{2pc}
\noindent{\it Keywords}: Al/Al2O3, metal-oxide interface, electron tunnelling, tunnel barrier, band offset, thin film, MIM junction 

\maketitle
\ioptwocol

\section{Introduction}
Regardless of numerous experimental and theoretical investigations of metal-oxide interfaces, very little is known about the effect of the interfacial geometry on electron transport. The issue is of paramount importance particularly for metal-insulator-metal (MIM) tunnel junctions where a thin oxide layer creates for electrons a potential barrier between two metals \cite{Ville2011,Martinis2014}. In particular, despite the vast popularity of  Al/Al$_2$O$_3$/Al junctions in modern nanoelectronics and continuous interest in their novel applications, for example, in qubits and Superconducting Quantum Interference Devices (SQUIDs) \cite{Al_AlOX_Al_QUBIT_2012,Al_AlOX_Al_QUBIT_2006,Al_AlOX_Al_QUBIT_2004}, single-electron transistors (SETs) \cite{Nakamura_SET_1996,Al_AlOX_Al_SET_1998}, energy storage \cite{Al_AlOX_Al_capacitor_2014}, infrared sensors \cite{IRsensor2015}, to the best of our knowledge, not a single theoretical study about the influence of the interface geometry on electron tunnelling has been performed for these systems. The existing theoretical studies, which are based on fitting the calculated data to the measured current-voltage characteristics ($I$-$V$), usually employ the standard Simmons model for calculating current densities \cite{Simmons1963}. The model suggests an analytic expression for the current density applicable to any shape of the barrier given that the average barrier height and width are known or represent fit parameters. Following the general model, current-voltage relationships are also deduced for the rectangular barrier including image force effects and considering various voltage ranges.
 
Usually, the thickness of the realistic insulating layer varies throughout the junction. Therefore, the width of the tunnel barrier (which may differ from the physical thickness of the insulator) also varies. This is of substantial importance because the conductance of the tunnel junction is known to depend exponentially on the barrier width \cite{Fisher1961}. Non-uniformity of the order of even one atomic layer may cause significant changes in the electron transport properties of the device \cite{Ville2011}. To address the non-uniformity in  the thickness (and in the oxidation level) of the insulating layer, a double-layer barrier model has been introduced in \cite{Arakawa2006}. In this model, the barrier heights and thicknesses are evaluated by fitting the numerically calculated tunnel probabilities to experimental data assuming a rectangular potential and employing Simmons model for the current density. 

The second primary parameter defining the tunnelling properties, \textit{i.e.}, the height of the potential barrier in the junction, may be affected by the metal states penetrating into the oxide gap. Thereby, additional energy levels are introduced for the tunnelling electrons, similarly to the metal-semiconductor junctions \cite{Tersoff1984}. To study the effect of such evanescent states on tunnel transport, the transfer matrix method  \cite{TsuEsaki1973} was applied in \cite{Jung2009} together with the Simmons model to calculate $I$-$V$ curves for rectangular and side-modified rectangular barriers. A similar method was employed in \cite{JungDFT} with an image-force-modified tunnel barrier profile to study the impact of the substrate on electron transport in Al/Al$_2$O$_3$/metal junctions. Based on the Simmons model, extensive analyses of experimental data were presented in \cite{Gloos2003} and it was concluded that there is a strong correlation between the barrier height and the thickness. Additionally, a discrete distribution of barrier thicknesses with the peak-to-peak distances corresponding to a single oxygen layer was observed.
 
In all the above-listed studies, the main approach has been the assumption of a certain shape of the barrier, mainly rectangular with possible modifications, and the fitting of the calculated $I$-$V$ data to the experimentally measured one in order to find the height and the width of the barrier. Although these studies provide valuable insight into the average barrier or transport properties, they lack the atomic-scale characterization. On the other hand, theoretical studies which are based on atomic modelling involve a single model geometry, leaving aside all the possible interface effects \cite{JungDFT,Bokes_conductance}. The goal of this work is to establish a link between the structural properties of the interface and the corresponding barrier parameters (\textit{i.e.}, the height and the width). Since in theory, there is an uncountable number of options to interface the two materials, it is impossible to examine all the potential configurations. Therefore, we study by first-principles electronic structure calculations six representative geometries of the Al/Al$_2$O$_3$ interface to extract the barrier parameters for electron tunnelling. The chosen structures differ in stacking sequences and oxide terminations. On the basis of our first-principles results and the experimental data, we demonstrate in this paper that both the barrier height as well as the thickness are highly dependent on the atomic arrangement and the stoichiometry at the interface. In addition, by fitting the semiclassically calculated conductance to the experimental $G$-$V$ curve and comparing the results to those obtained from the density-functional theory (DFT), we predict the average barrier parameters and the most expected geometry of the junction measured in our experiment.

\section{Experimental methods}
\begin{figure}[h!t]
\begin{center}
\includegraphics[width=0.45\textwidth]{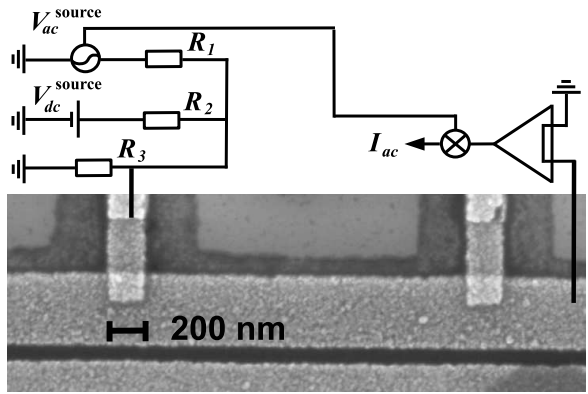}
\caption{SEM of two tunnel junctions together with a schematic of the experimental setup.}
\label{fig:1}
\end{center}
\end{figure}

In the following, we describe the fabrication and measurement techniques. Single tunnel junctions are made by electron beam lithography and shadow evaporation technique \cite{Fulton1987}. Several Al-Al$_x$O$_y$-Al junctions were formed between the fingers and the common electrode. The thickness of all the fingers and the common electrode is 25 nm. The tunnel barrier (AlOx) of the junction  was formed in-situ by thermal oxidation before the deposition of the second layer of Al. A close-up of two tunnel junctions is shown in the scanning-electron micrograph (SEM) together with the experimental setup in figure ~\ref{fig:1}.

We measure the differential conductance $G$ in a liquid helium dipstick at $T_{bath}$ = 4.2~K with a standard lock-in technique. The measurement setup shown in figure~\ref{fig:1} consists of dc and ac voltage sources, low noise current preamplifier (DL instruments 1211), and a lock-in amplifier (SR 830). We obtain the differential conductance as $G =  I_{ac}/ V_{ac}$, where $ I_{ac}$ is the measured ac current through the junction and $V_{ac} = V^{source}_{ac}R_{3}/(R_{1} + R_{3})$ the applied ac bias voltage calculated according to the voltage division. The dc voltage is $V_{dc} = V^{source}_{dc}R_{3}/(R_{2} + R_{3})$.

\begin{figure}[h!t]
\begin{center}
\includegraphics[width=0.4\textwidth]{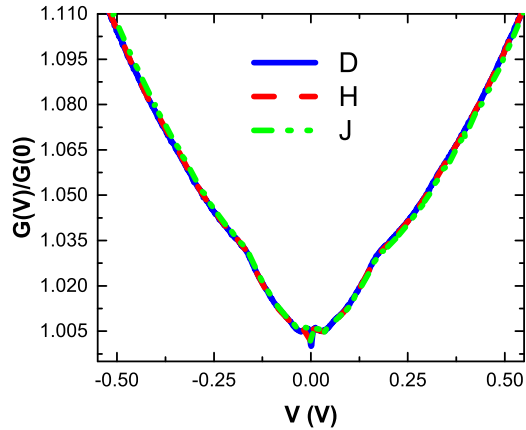}
\caption{(color online) Measured differential conductance vs voltage for three single junctions.}
\label{fig:2}
\end{center}
\end{figure}

The measured differential conductance curves of three junctions are shown in figure~\ref{fig:2} as solid blue (sample D), dashed red (sample H), and dash-dotted green (sample J). The differential conductance has a parabolic shape and a Coulomb blockade dip at zero bias voltage. Parabolic dependence is also the lowest-order result from Simmons model for tunnelling through a
rectangular barrier \cite{Simmons1963}. The zero-bias conductance is determined through the parabolic fit with the $I$-$V$ curve and is found to be 1802.17 $\mu S/ \mu m^2$,  1794.5 $\mu S/ \mu m^2$ and 1765.5 $\mu S/ \mu m^2$ for the three junctions. Coulomb blockade dip is ignored in the fitting.

\section{Theoretical methods}
\subsection{Modelling Al/Al$_2$O$_3$ interfaces}

\begin{figure}
\begin{center}
\includegraphics[width=0.5\textwidth, trim={1.0cm 2.8cm 0 0.1cm}]{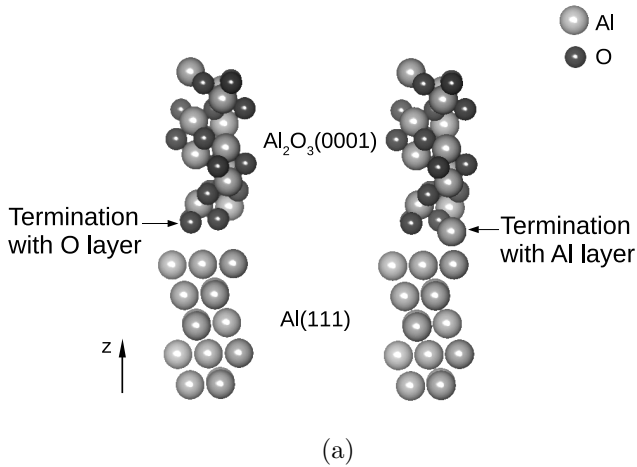}
\caption*{(a)}
\label{term}
\end{center}

\begin{center}
\includegraphics[width=0.5\textwidth, trim={0.5cm 4.8cm 0 1.5cm}]{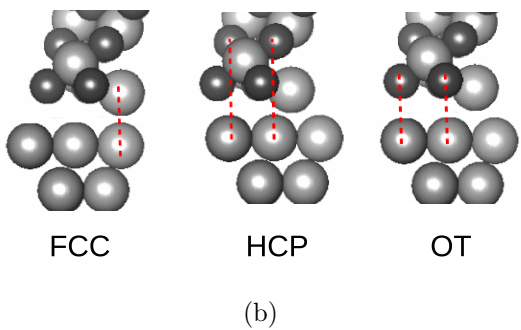}
\caption*{(b)}
\label{stack}
\caption{Illustration of the modelled structures. (a) The two possibilities of the oxide termination. O and Al layer-terminated Al$_2$O$_3$ are interfaced with the Al(111) substrate. (b) The stacking sequences, an example of the Al-terminated interface. FCC: Al surface atoms of the metal and the oxide sit on top of each other, HCP: Al atoms are placed along the second O layer of the oxide, OT: Al atoms from the metal sit on top of the first O layer of the oxide. The dashed lines connect the corresponding layers.}
\label{geometry}
\end{center}
\end{figure}

In our first-principles calculations, the model junction consists of a 5-layer substrate for the Al(111) surface and either single Al layer- or O layer-terminated oxide (figure \ref{geometry}). The oxide is represented as the crystalline Al$_2$O$_3$ having hexagonal unit cell and the (0001) surface parallel to the interface. For each termination we examine interface structures with three different stacking sequences: face-centred cubic (FCC), hexagonal closed packed (HCP) and octahedral (OT) (figure \ref{geometry}(b)). Thus, in total, we investigate six different structures. Throughout the paper we label the interfaces as "1Al FCC", etc., which read as "single Al layer-terminated oxide with FCC stacking sequence", etc. For example, the structures presented in figure \ref{geometry}(a) correspond to the O FCC and 1Al FCC interfaces. To extract barrier parameters, we first relax the chosen structures using DFT with the GPAW code \cite{GPAWrev,GPAW}. We carry out calculations with the 4x4x1 Monkhorst-Pack k-point grid and the PBE exchange-correlation functional \cite{PBE}. The simulation box includes 5 \AA ~vacuum on each side of the slab along the $z$ axis. To avoid the artificial electric field due to asymmetry, we apply the dipole correction to the electrostatic potential \cite{Dipole_correction_1,Dipole_correction_2} perpendicularly to the interface, as implemented in GPAW. Further details of the calculation setup and the procedure for selecting the interfaces, as well as a thorough discussion on their mechanical and electronic properties will be provided in a separate paper.

\subsection{Extraction of barrier parameters}

\begin{figure}
\begin{center}
\includegraphics[width=0.5\textwidth, trim={0.0cm 0.0cm 0.0cm 0.0cm}]{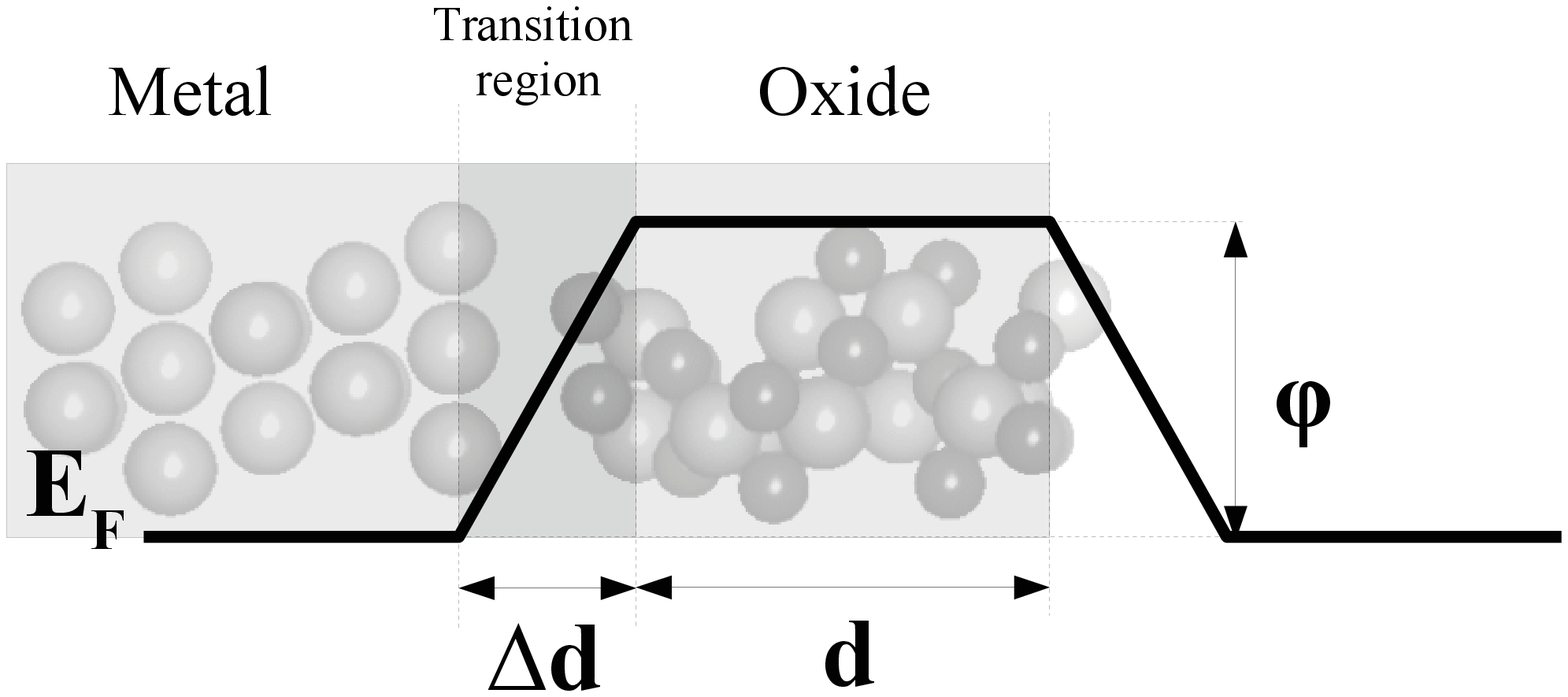} 
\caption{Trapezoidal barrier model. $E_F$ is the Fermi level of the metal, $\Delta d$ is the width of the transition region at the metal-oxide interface, $d$ is the barrier width corresponding to the smaller base of the trapezoid, and $\phi$ is the barrier height.}
\label{profile}
\end{center}
\end{figure}

The simplest model for the profile of the tunnel barrier is a one-dimensional rectangular potential wall corresponding to an abrupt transition between the metal and the oxide. However, the importance of accounting for a transition region (along with an electron effective mass for the oxide) has been pointed out in \cite{Bokes_conductance}. Hence, we start by assuming a trapezoidal shape for the potential. To construct barrier profiles for our systems,  we need to know the barrier heights $\phi$, transition region widths $\Delta d$ and the widths $d$ of the barriers for each geometry (figure \ref{profile}). In the following section, we describe the details for determining the three parameters.

\subsubsection{Barrier height $\phi$}

\begin{figure}
\begin{center}
\includegraphics[width=0.5\textwidth, trim={0.5cm 0.0cm 0.0cm 0.0cm}]{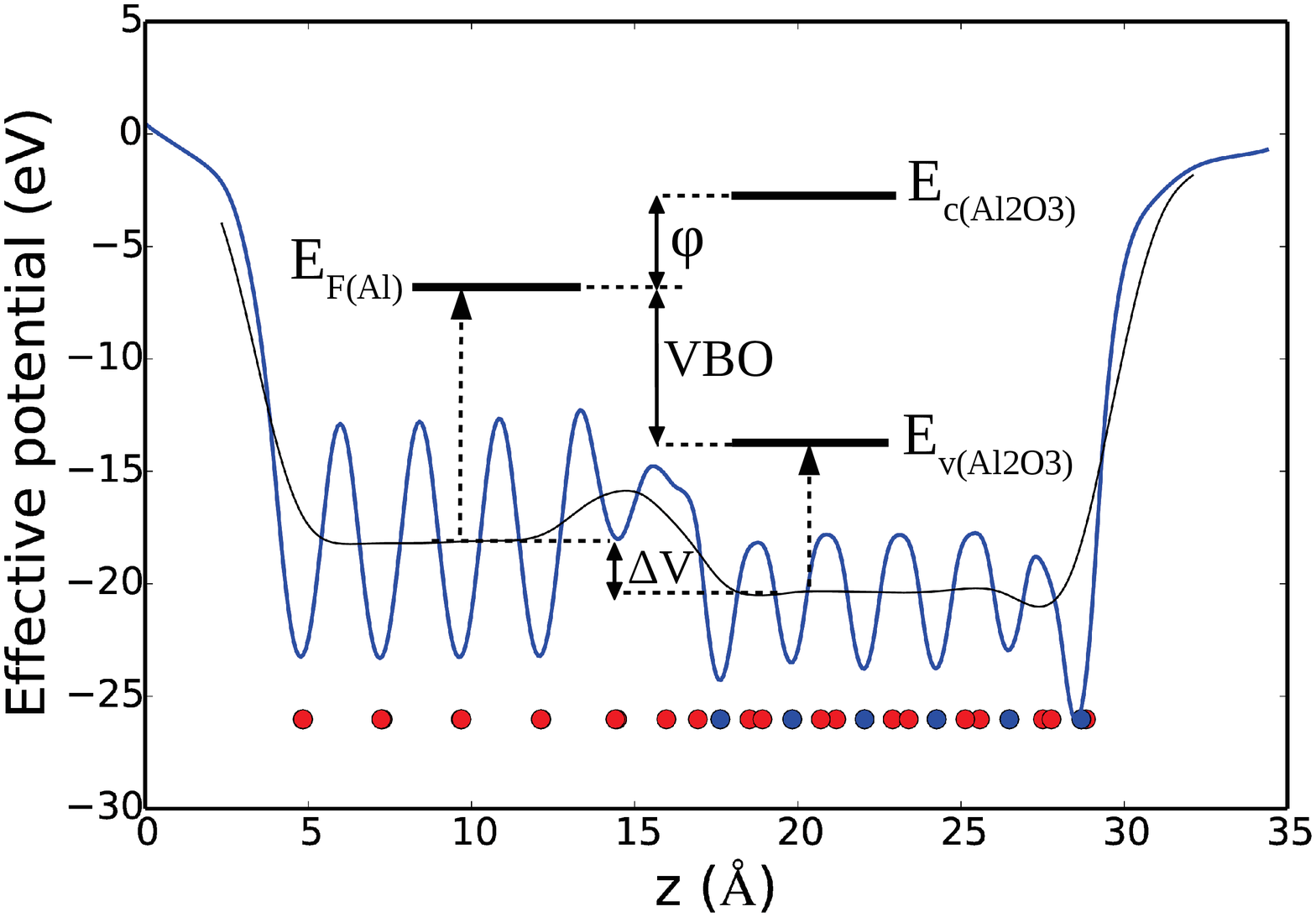}
\caption{(color online) Schematics of the band alignment. The red and blue circles represent Al and Al$_2$O$_3$ layers, respectively. The thick continuous blue line is the plane-averaged effective potential along the direction perpendicular to the interface. The thin continuous black line shows the macroscopically averaged plane-averaged effective potential. $\phi$ is the tunnel barrier height. The fine-dashed arrow on the left-hand side shows the difference between the Fermi level  of the \textit{bulk} Al and its average effective potential. Similarly, the fine-dashed arrow on the right-hand side shows the difference between the VBM of the \textit{bulk} Al$_2$O$_3$ with respect to its average effective potential. E$_\textrm{{c(Al$_2$O$_3$)}}$ is the CBM of the oxide. $\Delta V$ is the potential shift. VBO is equivalent to the sum of two quantities, the band-structure term ($E_V$) and the potential line-up ($\Delta V$) (see (\ref{vbo})).}
\label{alignment}
\end{center}
\end{figure}

The tunnel barrier height is estimated according to the Schottky model \cite{Schottky}:

\begin{equation}
\phi=W-X,
\label{schottky_height}
\end{equation}
where $W$ is the work function of the metal, \textit{i.e.}, the difference between the Fermi level ($E_F$) of the metal and the vacuum level, and $X$ is the electron affinity of the oxide, \textit{i.e.}, the difference between the conduction band minimum (CBM) of the oxide and the vacuum level. Since the definition of $\phi$ is based solely on the properties of the two isolated bulk materials, it neglects the effects of the interface formed between them. In particular, due to the charge transfer across the interface, a dipole barrier is exerted in this region which may be different for different interface geometries and chemical compositions. On the other hand, because of the long-range character of the Coulomb interaction the band alignment between the adjacent metal and the oxide is determined, in addition to the bulk properties, by the electronic distribution at the interface. As a consequence, to apply (\ref{schottky_height}) to the interface, one needs to know the position of the oxide CBM inside the junction. Since PBE is known to underestimate band gaps, instead of using the theoretical value for CBM, it is more convenient to first find the valence band minimum (VBM) of the oxide. Next, CBM can be obtained by adding the experimental value of the oxide band gap ($E_{g(exp.)}$) to the VBM found from DFT. For the experimental band gap we use a value of 8.8 eV \cite{French1990}.  Thus (\ref{schottky_height}) is equivalent to:

\begin{equation}
\phi=(VBM+E_{g(exp.)})-E_F,
\label{phi1}
\end{equation}

or,
\begin{equation}
\phi=VBO+E_{g(exp.)},
\label{phi2}
\end{equation}
where VBO is the valence band offset relative to the Al Fermi level. Thus, the problem maps to finding the VBO for the interface. For this purpose we use the \textit{macroscopic average} technique which was first developed for finding valence band offsets in lattice-matched metal-semiconductor heterojunctions and was further extended to the lattice-mismatched case \cite{Lattice_matched,Lattice_mismatched,band_method_review}. According to the method the valence band offset reads as: 

\begin{equation}
VBO=E_V+\Delta V,
\label{vbo}
\end{equation}
where $E_V$ is the band-structure term and $\Delta V$ is the potential line-up. $E_V$ is defined as the difference between the VBM of the oxide and the Fermi level of the metal, both measured with respect to the macroscopic average electrostatic potentials in the corresponding bulk materials. Therefore, $E_V$ is independent of the interface and is defined by the bulk band structure, only. In contrast, $\Delta V$ is an interface-dependent term and represents the difference between the macroscopic average electrostatic potentials in the bulk-like regions of the interfaced materials. Figure \ref{alignment} illustrates the calculation procedure for the barrier heights on the example of 1Al FCC interface. 

\begin{table*}
\caption{\label{offsets} Valence band offsets (VBO) and barrier heights $\phi$ for the different interface geometries.} 
\lineup
\noindent\begin{tabular}{@{}*{7}{l}}
\br  
\textbf{Geometry} & 1Al FCC & 1Al HCP & 1Al OT & O FCC & O HCP & O OT \cr
\mr 
\textbf{VBO(eV)} & $-7.65$ & $-5.65$ & $-6.15$ & $-7.55$ & $-7.55$ & $-7.05$ \cr
\boldmath{$\phi$}\textbf{(eV)} & \m1.15 & \m3.15 & \m2.65 & \m1.25 & \m1.25 & \m1.75 \cr
\br 
\end{tabular}
\end{table*}

Since the junction is lattice-mismatched and, correspondingly, the periodicity of $V(z)$ is different in the metal and in the oxide bulk regions, the macroscopic averaging of the potential is done twice successively: first with the window size matched with the periodicity of the bulk Al, next - with the periodicity of the bulk Al$_2$O$_3$. 

The Schottky barrier heights obtained for various interfaces are summarized in table \ref{offsets}. Our results show that the geometry of the interface highly influences the tunnelling properties as the difference in the barrier heights can be as large as 2 eV, against the average value of 1.87 eV. On the other hand, it is gratifying to note that the barrier heights derived based on experimental data are reported to be approximately 2 eV \cite{JungDFT}. The barrier heights for the Al-terminated interfaces are, on average, higher than those of the O-terminated interfaces, with the mean heights being 2.32 eV and 1.42 eV, respectively. This gives rise to an about 1 eV difference between the average heights of the two different terminations.

\subsubsection{Transition region width $\Delta d$}

\begin{figure}
\begin{center}
\includegraphics[width=0.5\textwidth, trim={0.0cm 0.0 0.0cm 0.0}]{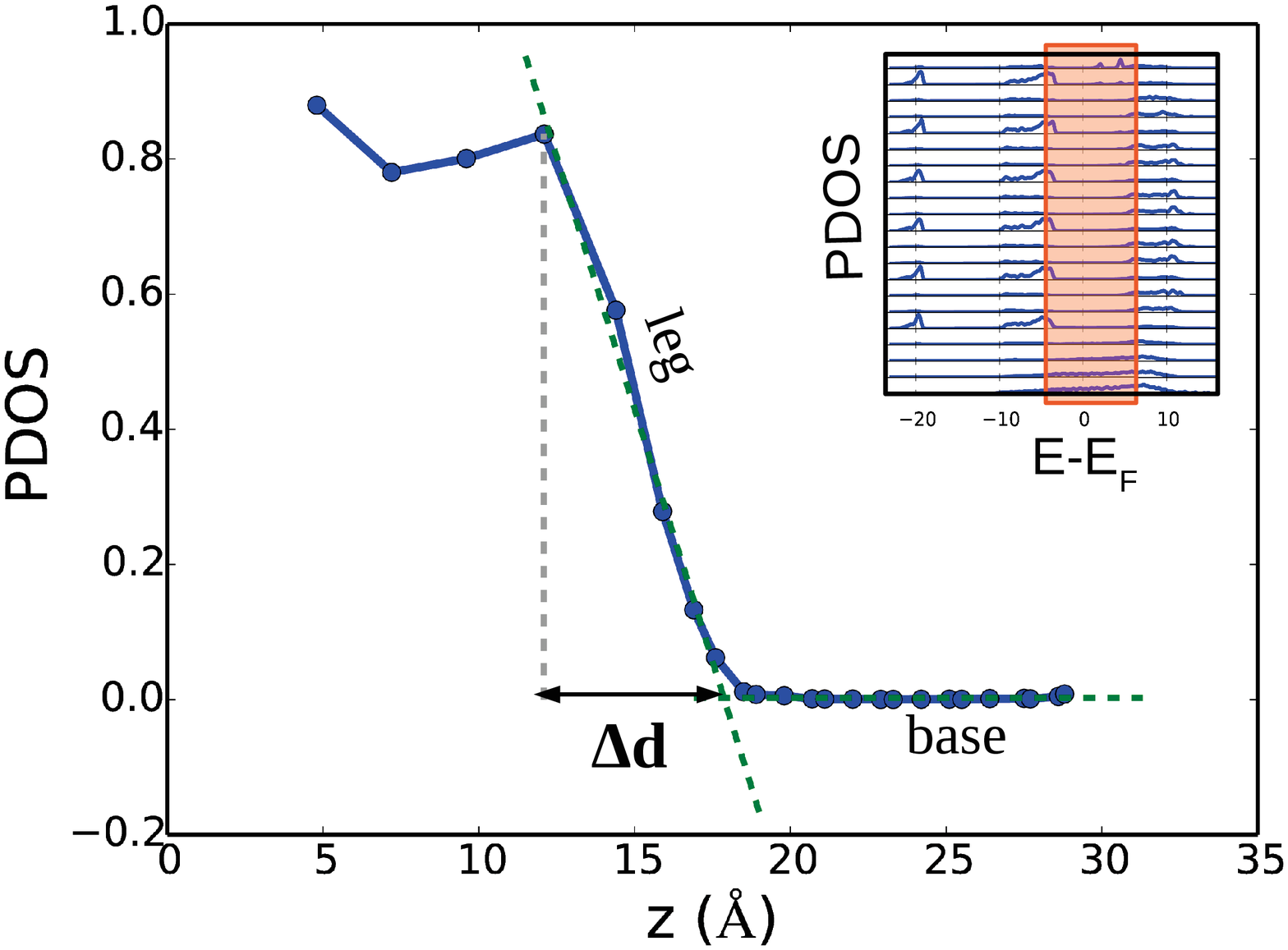} 
\caption{(color online) Layer-projected density of states as a function of the $z$ coordinate. The fit of the trapezoid is shown with the thick dashed green lines. $\Delta d$ is the estimated width of the transition region. The inset illustrates the averaging window for the PDOS.}
\label{pdosz}
\end{center}
\end{figure}

Assuming that the local density of states follows the shape of the barrier along the coordinate perpendicular to the interface \cite{JungDFT, Bokes_conductance}, we extract the width of the transition region between the metal and the oxide from the projected density of states (PDOS) for atomic layers at the interface. For this reason, we perform linear fittings for the base and the leg of the trapezoid separately (figure \ref{pdosz}). The layer-projected densities of states are plotted as a function of the $z$ coordinate, and averaged over an energy window centred at the Fermi level (inset in the upper right corner).

The obtained results are summarized in table \ref{deld}. Interestingly, the transition widths of  both the Al- and O-terminated interfaces follow the same trend: FCC, HCP, and OT in the order of decreasing width. The significant magnitude of $\Delta d$ indicates a pronounced deviation from the square barrier assumption, where an abrupt transition between the metal and the oxide is assumed. It must be noted that the presence of the transition region lowers the average height of the zero-bias barrier by a factor of $2/3$ compared to the rectangular barrier, since $\left\langle \phi_{rect.} \right\rangle=\phi_{rect.}$, and $\left\langle \phi_{trapz.} \right\rangle=(2/3) \phi_{trapz.}$ at $V=0$. This suggests that the barrier heights determined by fitting the rectangular barrier models (such as the widely used Simmons model \cite{Simmons1963}) with the experimental $I$-$V$ curves should be expected, in fact, to be higher by a factor of $3/2$ for symmetrical barriers.

\begin{table*}
\caption{Transition region widths ($\Delta d$) for the different geometries extracted from the layer-PDOS.} \label{deld}

\lineup
\noindent\begin{tabular}{@{}*{7}{l}}
\br 
\textbf{Geometry} & 1Al FCC & 1Al HCP & 1Al OT & O FCC & O HCP & O OT \cr
\mr
\boldmath{$\Delta d$}\textbf{(\AA)} & 5.83 & 2.78 & 2.62 & 3.34 & 3.15 & 2.05 \cr
\br
\end{tabular}
\end{table*}

\subsubsection{Barrier width $d$}

The last parameter of our tunnel barrier model is the width $d$. To analyse the experimental structures, we assume that the interface studied in the experiment is mainly composed of one of the modelled structures. For each case we fit the width with the measured zero-bias conductance. According to reference \cite{Bokes_conductance} the conductance through our model barrier is:

\begin{equation}
G_{0(theor.)} \approx - \frac{e^{-F(E_F)}}{2\pi^2F^\prime(E_F)},
\end{equation}

where

\begin{equation}
F(E_F)=2\sqrt{2m^*\phi}\left(d+\frac{2}{3}\Delta d\right)
\end{equation}

and

\begin{equation}
F^\prime (E_F)=- \frac{2}{\sqrt{2m^*\phi}}\left(d+2\Delta d\right).
\end{equation}

The above expression has been derived following the Simmons model adopted for the trapezoidal barrier profile and taking into account the contribution of the electrons to tunnelling only from the states close to the Fermi energy. We calculate the effective electron mass $m^*$ from the DFT band structure of the bulk oxide. Our obtained value, 0.38 $m_e$, is in a perfect agreement with earlier experimental and theoretical works \cite{Perevalov2007,Medvedeva2007}. The obtained widths are displayed in table \ref{d}.

\begin{table*}
\caption{Barrier widths $d$ found by fitting the theoretical zero bias conductance to the experimentally measured one.} \label{d}
\begin{tabular}{@{}*{7}{l}}

\br
\textbf{Geometry} & 1Al FCC & 1Al HCP & 1A OT & O FCC & O HCP & O OT \cr
\mr
\boldmath{$d$}\textbf{(\AA)} & 8.89 & 6.77 & 7.54 & 10.34 & 10.48 & 9.70 \cr
\br

\end{tabular}
\end{table*}

\begin{figure}[h]
\begin{center}
\includegraphics[width=0.45\textwidth]{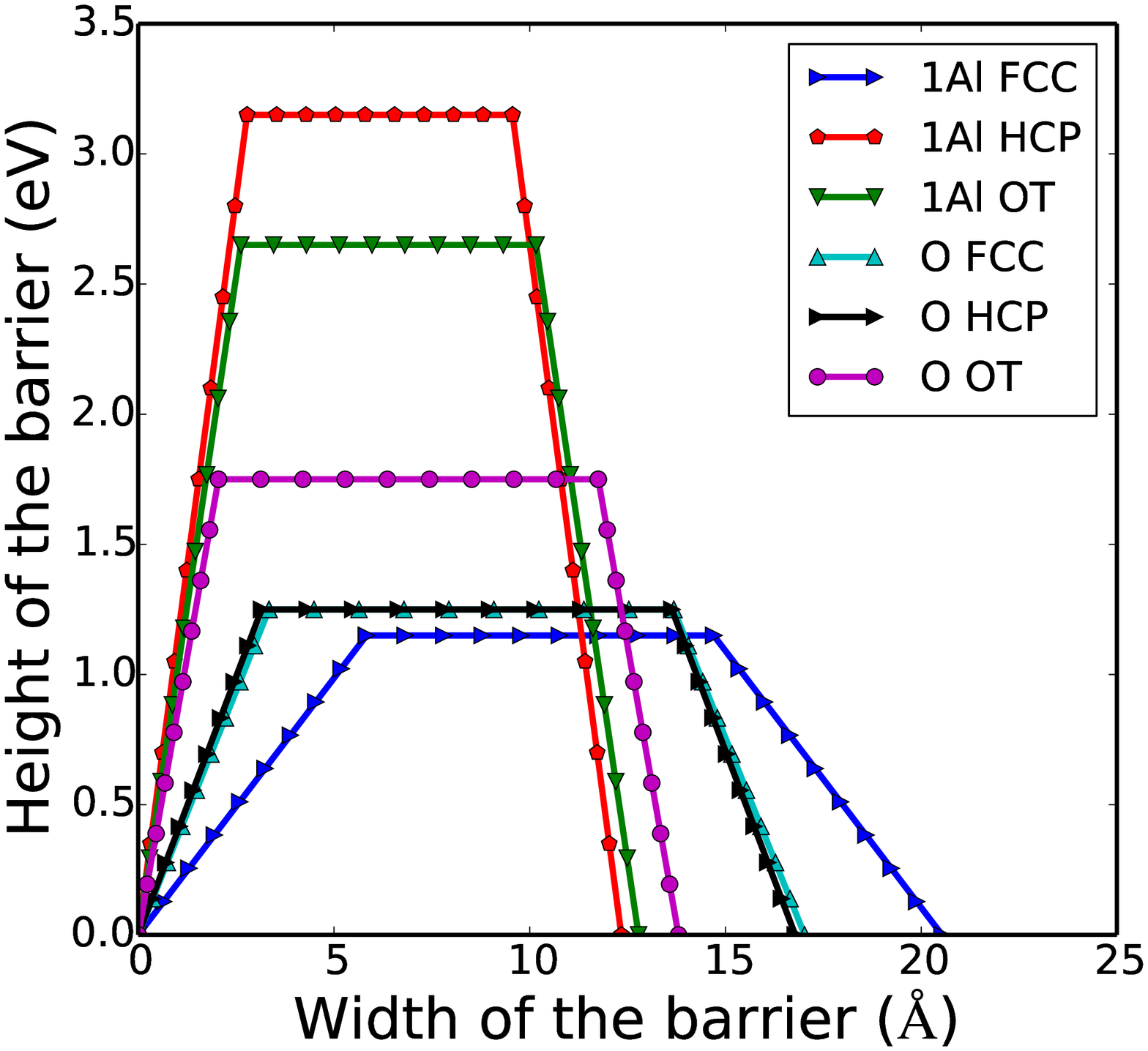}
\caption{(color online) Predicted barrier profiles.\label{barriers}}
\end{center}
\end{figure}

Our results demonstrate that O-terminated interfaces exhibit wider barriers compared to the Al-terminated counterparts, with average values of 7.73 \AA~ and 10.17 \AA, respectively. Based on the data in tables \ref{offsets}-\ref{d}, resulted barrier profiles are visualized in figure \ref{barriers}. 

\section{Analysis of the experimental data}

Since the widths were extracted by fitting to a single point, the zero-bias conductance, this way, we obtain the picture only about the relative widths of the different barriers with respect to each other. In addition, the electron effective mass of the oxide, which was fixed at the bulk value, might differ for each interface. Furthermore, we assumed that the interface measured in the experiment was composed of only one of the discussed interfaces, which might not be the case in the real junction. Therefore, to explain the experimental data, we extend our analyses beyond the zero bias. We make an average evaluation of the barrier parameters, and of the effective mass, using the WKB approximation by fitting the conductance to the experimental $G$-$V$ curve. $\phi$, $\Delta d$, $d$ and $m^*$ are left as free parameters. The tunnel probability along the $z$ direction within the WKB approximation is calculated with the following expression:

\begin{equation}
P(E_z)=\exp\left\{ -\frac{4\pi}{h} \int_{z_1}^{z_2} \sqrt{2m^*\left[\phi(z,V)-E_z\right]} dz \right\}.
\end{equation} ~

\begin{figure}[]
\begin{center}
\includegraphics[width=0.4\textwidth]{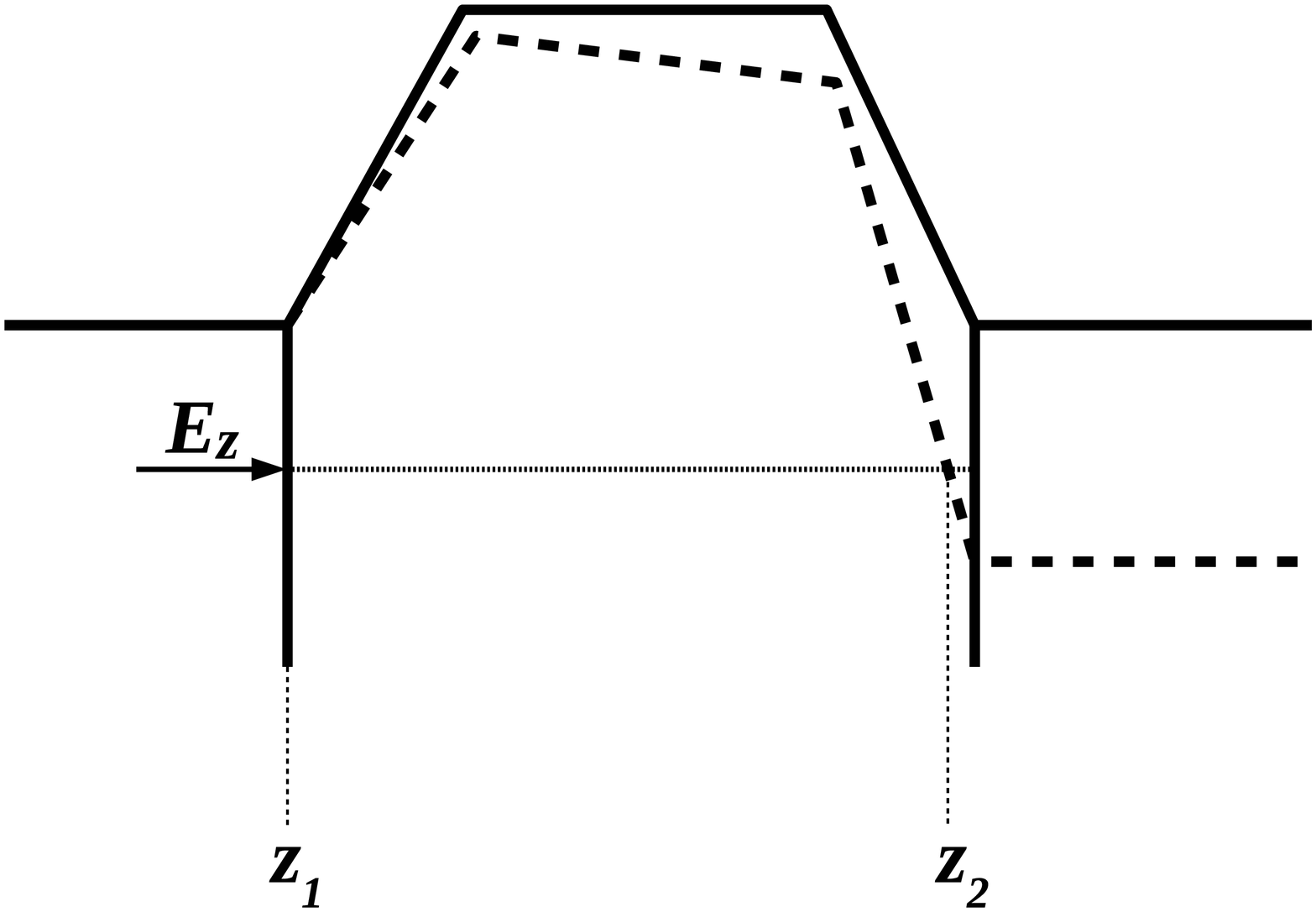}
\caption{Schematic representation of the trapezoidal barrier tilt under bias. The solid line represents the barrier profile at V=0. The dashed line shows the distorted barrier due to the applied voltage V. $E_z$ is the energy of the incident electron and $z_1$ and $z_2$ are the classical turning points at the energy $E_z$.}
\label{tilt}
\end{center}
\end{figure}

And the ensuing current density is:

\begin{eqnarray}
\fl
j =\frac{4\pi m^*q}{h^3}kT\int_0^\infty{P(E_z)} \\
\nonumber\times \ln\left\{ \frac{1+\exp\left[(E_F-E_z)/kT\right]}{1+\exp\left[(E_F-Ez-qV)/kT\right]}\right\}dE_z,
\end{eqnarray} ~

where $m^*$ is the effective electron mass, $E_z$ is the energy of the incident electron, and $z_1$ and
$z_2$ are the classical turning points at the given energy $E_z$ (figure \ref{tilt}).

As it is shown in figure \ref{fig:2}, the experimental curve exhibits a kink at about 0.17 V. The behaviour above this point has not been explained consistently in the existing studies, although, it has been suggested earlier to be the result of electron-phonon scattering \cite{Lau1981}. Therefore, we consider only the bias range below 0.17 V as the WKB approximation does not account for inelastic effects. We apply a tilt in a form of a linear potential to the trapezoid as a modification caused by the bias voltage (figure \ref{tilt}). Unlike for a rectangle, the width of the trapezoid is now reduced due to the tilt and it is dependent on $E_z$. The fitting, which reproduces essentially the parabolic $G$-$V$ dependence,
results in a barrier height of 2.56 eV, a transition region width of 1.70 \AA, and a barrier width of 8.43 \AA. The obtained barrier height is significantly higher than the barrier height corresponding to any O-terminated interface found from the DFT calculations. In contrast, it is closest to that of the 1Al OT configuration with the error of only 3$\%$. The second best agreement between the \textit{ab-initio} and fitted values is found for the 1Al HCP configuration with the error of 18$\%$. This indicates that the measured junction (mainly) consists of the Al-terminated interfaces. However, this reasoning cannot totally exclude the presence of O-terminated interfaces. The transition region width is underestimated compared to those obtained using DFT, with the biggest errors for the 1Al FCC, O FCC and O HCP geometries. Concentrating on the Al-terminated cases, the best agreement is found, again, for the 1Al OT and 1Al HCP interfaces, what further supports our previous prediction. The fitted and the DFT values of $\Delta d$ agree by 35$\%$ and 39$\%$ for 1Al OT and 1Al HCP, respectively. Taking into account that in our first-principles calculations our model crystalline geometries are idealistic approximations to the real junction geometry, which in fact, is amorphous, we consider the agreement between the two methods rewarding. 
The electron effective mass derived from fitting, 0.29 $m_e$, is smaller than the calculated bulk value. 

\section{Summary}

In conclusion, by combining first-principles density-functional and classical methods, we have extracted the tunnel barrier profiles for six possible interface geometries of the Al/Al$_2$O$_3$ interface. Variations in the obtained barrier heights, the barrier widths and the widths of the transition regions demonstrate that the electron transport properties of the junctions are highly sensitive specifically to the geometry of the interface. We obtained that the O-terminated interfaces exhibit on average 0.9 eV lower and 2.44 \AA~ wider barriers compared to the Al-terminated ones. The resulted transition region widths suggest that the real barriers are better represented by the trapezoid. In addition, by fitting the semiclassical model to the experimental $G$-$V$ curve, we have obtained information about the average barrier properties in a real junction. When compared with our DFT-based results the fitting predicts that the interfaces are Al-terminated and are best described by the octahedral stacking. The gained information is important for understanding electron transport through tunnel barriers on the atomistic level.

\ack
We acknowledge the availability of the facilities and technical support by Otaniemi research infrastructure for Micro and Nanotechnologies (Otanano). We acknowledge financial support from the European Community FP7 Marie Curie Initial Training Networks Action (ITN) Q-NET 264034  and the Academy of Finland through its CoE programme (projects no. 251748, 284621, 250280 and 284594). We acknowledge the computational resources provided by the Aalto Science-IT project. We have benefitted from fruitful discussions with D. Averin.

\section*{References}

\bibliographystyle{unsrt} 
\bibliography{article} 

\end{document}